\documentclass[12pt]{article}
\usepackage{html}
\usepackage{epsf}
\usepackage{graphicx}
\usepackage{amssymb}
\begin{document}
\title{MATTERS OF GRAVITY, The newsletter of the APS Topical Group on 
Gravitation}
\begin{center}
{ \Large {\bf MATTERS OF GRAVITY}}\\ 
\bigskip
\hrule
\medskip
{The newsletter of the Topical Group on Gravitation of the American Physical 
Society}\\
\medskip
{\bf Number 41 \hfill Winter 2013}
\end{center}
\begin{flushleft}
\tableofcontents
\vfill\eject
\section*{\noindent  Editor\hfill}
David Garfinkle\\
\smallskip
Department of Physics
Oakland University
Rochester, MI 48309\\
Phone: (248) 370-3411\\
Internet: 
\htmladdnormallink{\protect {\tt{garfinkl-at-oakland.edu}}}
{mailto:garfinkl@oakland.edu}\\
WWW: \htmladdnormallink
{\protect {\tt{http://www.oakland.edu/?id=10223\&sid=249\#garfinkle}}}
{http://www.oakland.edu/?id=10223&sid=249\#garfinkle}\\

\section*{\noindent  Associate Editor\hfill}
Greg Comer\\
\smallskip
Department of Physics and Center for Fluids at All Scales,\\
St. Louis University,
St. Louis, MO 63103\\
Phone: (314) 977-8432\\
Internet:
\htmladdnormallink{\protect {\tt{comergl-at-slu.edu}}}
{mailto:comergl@slu.edu}\\
WWW: \htmladdnormallink{\protect {\tt{http://www.slu.edu/colleges/AS/physics/profs/comer.html}}}
{http://www.slu.edu//colleges/AS/physics/profs/comer.html}\\
\bigskip
\hfill ISSN: 1527-3431

\bigskip

DISCLAIMER: The opinions expressed in the articles of this newsletter represent
the views of the authors and are not necessarily the views of APS.
The articles in this newsletter are not peer reviewed.

\begin{rawhtml}
<P>
<BR><HR><P>
\end{rawhtml}
\end{flushleft}
\pagebreak
\section*{Editorial}

The next newsletter is due September 1st.  This and all subsequent
issues will be available on the web at
\htmladdnormallink 
{\protect {\tt {https://files.oakland.edu/users/garfinkl/web/mog/}}}
{https://files.oakland.edu/users/garfinkl/web/mog/} 
All issues before number {\bf 28} are available at
\htmladdnormallink {\protect {\tt {http://www.phys.lsu.edu/mog}}}
{http://www.phys.lsu.edu/mog}

Any ideas for topics
that should be covered by the newsletter, should be emailed to me, or 
Greg Comer, or
the relevant correspondent.  Any comments/questions/complaints
about the newsletter should be emailed to me.

A hardcopy of the newsletter is distributed free of charge to the
members of the APS Topical Group on Gravitation upon request (the
default distribution form is via the web) to the secretary of the
Topical Group.  It is considered a lack of etiquette to ask me to mail
you hard copies of the newsletter unless you have exhausted all your
resources to get your copy otherwise.

\hfill David Garfinkle 

\bigbreak

\vspace{-0.8cm}
\parskip=0pt
\section*{Correspondents of Matters of Gravity}
\begin{itemize}
\setlength{\itemsep}{-5pt}
\setlength{\parsep}{0pt}
\item Daniel Holz: Relativistic Astrophysics,
\item Bei-Lok Hu: Quantum Cosmology and Related Topics
\item Veronika Hubeny: String Theory
\item Pedro Marronetti: News from NSF
\item Luis Lehner: Numerical Relativity
\item Jim Isenberg: Mathematical Relativity
\item Katherine Freese: Cosmology
\item Lee Smolin: Quantum Gravity
\item Cliff Will: Confrontation of Theory with Experiment
\item Peter Bender: Space Experiments
\item Jens Gundlach: Laboratory Experiments
\item Warren Johnson: Resonant Mass Gravitational Wave Detectors
\item David Shoemaker: LIGO Project
\item Stan Whitcomb: Gravitational Wave detection
\item Peter Saulson and Jorge Pullin: former editors, correspondents at large.
\end{itemize}
\section*{Topical Group in Gravitation (GGR) Authorities}
Chair: Manuela Campanelli; Chair-Elect: 
Daniel Holz; Vice-Chair: Beverly Berger. 
Secretary-Treasurer: James Isenberg; Past Chair:  Patrick Brady;
Members-at-large:
Laura Cadonati, Luis Lehner,
Michael Landry, Nicolas Yunes,
Curt Cutler, Christian Ott,
Jennifer Driggers, Benjamin Farr.
\parskip=10pt

\vfill
\eject

\vfill\eject

\section*{\centerline
{we hear that \dots}}
\addtocontents{toc}{\protect\medskip}
\addtocontents{toc}{\bf GGR News:}
\addcontentsline{toc}{subsubsection}{
\it we hear that \dots , by David Garfinkle}
\parskip=3pt
\begin{center}
David Garfinkle, Oakland University
\htmladdnormallink{garfinkl-at-oakland.edu}
{mailto:garfinkl@oakland.edu}
\end{center}

Irwin Shapiro has received the Einstein Prize

Thomas Carruthers, Steven Giddings, Scott Hughes, Balasubramanian Iyer, Sergey Klimenko, Carlos Lousto, and Sheila Rowan
have been elected APS Fellows.

Hearty Congratulations!

\section*{\centerline
{100 years ago}}
\addtocontents{toc}{\protect\medskip}
\addcontentsline{toc}{subsubsection}{
\it 100 years ago, by David Garfinkle}
\parskip=3pt
\begin{center}
David Garfinkle, Oakland University
\htmladdnormallink{garfinkl-at-oakland.edu}
{mailto:garfinkl@oakland.edu}
\end{center}

In 1913 Einstein and Grossmann took the first steps towards a geometric theory of gravitation.  Their paper,
``{\it Entwurf einer verallgemeinerten Relativit\"{a}tstheorie und einer Theorie der Gravitation}'' [Outline
of a Generalized Theory of Relativity and of a Theory of Gravitation] 
published in {\it Z. Math. Phyzik.} {\bf 62}, 225 can be found in English translation at
\htmladdnormallink
{\protect {\tt{http://www.pitt.edu/\~jdnorton/teaching/GR\&Grav\_2007/pdf/Einstein\_Entwurf\_1913.pdf}}}
{http://www.pitt.edu/\~jdnorton/teaching/GR\&Grav\_2007/pdf/Einstein\_Entwurf\_1913.pdf}\\

\section*{\centerline
{GGR program at the APS meeting in Denver, CO}}
\addtocontents{toc}{\protect\medskip}
\addcontentsline{toc}{subsubsection}{
\it GGR program at the APS meeting in Denver, CO by David Garfinkle}
\parskip=3pt
\begin{center}
David Garfinkle, Oakland University
\htmladdnormallink{garfinkl-at-oakland.edu}
{mailto:garfinkl@oakland.edu}
\end{center}

We have an exicting GGR related program at the upcoming APS April meeting in Denver, CO.  Our Chair-Elect, Daniel Holz, did 
an excellent job of putting together this program.  At the APS meeting there will be several invited sessions of talks 
sponsored by the Topical Group in Gravitation (GGR).  

The invited sessions sponsored by GGR are as follows:\\

Relativistic Turbulence and MHD\\
Andrew MacFadyen, (TBA)\\
Bruno Giacomazzo, General Relativistic Magnetohydrodynamic Simulations of Compact Binary Mergers\\
Patrick Fragile, General Relativistic Radiation Magnetohydrodynamic Simulations of Black Hole Accretion\\

Recent Developments in Mathematical Relativity\\
Sergiu Klainerman, (TBA)\\
Mihalis Dafermos, (TBA)\\
Robert Wald, Dynamic and Thermodynamic Stability of Black Holes\\

Tidal Disruption Events\\
(joint with DAP)\\
Suvi Gezari, Observations of Tidal Disruptions by Black Holes\\
Enrico Ramirez-Ruiz, Simulations of Tidal Disruptions\\
Linda Strubbe, Predictions for Observational Signatures of the Tidal Disruption of Stars\\

Instrumentation for Current and Future Gravitational Wave Detectors\\
Brian O'Reilly, Status of Current Detectors\\
Lisa Barsotti, Beyond Advanced Gravitational Wave Detectors: Beating the Quantum Limit with Squeezed States of Light\\
Nicolas Smith-Lefebvre, Future Detectors II\\

Gravitational Wave Astrophysics\\
(joint with DAP)\\
Chris Fryer, Gravitational Waves: Probes of Stellar Collapse\\
Jocelyn Read, Learning About Dense Matter From Gravitational Waves\\
Will Farr, Gravitational Waves From Binaries and Dense Stellar Clusters\\

Black Hole Firewalls\\
Donald Marolf, Are There Surprising Quantum Gravity Effects Near the Horizons of Large Black Holes?\\
Raphael Bousso, (TBA)\\
Steve Giddings, (TBA)\\

Future Gravitational Wave Missions from Space\\
Paul McNamara, The LISA Pathfinder Mission\\
John Conklin, Gravitational Reference Sensor Technology Development at the University of Florida\\
Tyson Littenberg, A Stroll with eLISA Through the mHz Gravitational Wave Zoo\\
Neil Cornish, Observing Black Hole Mergers with Space Based Gravitational Wave Detectors\\
Kevin Kern, Multimessenger Astronomy: Modeling Gravitational and Electromagnetic Radiation from a Stellar Binary System\\

Einstein Prize Session\\
Irwin Shapiro, The Anatomy of a Test of General Relativity\\
Tom Murphy, Lunar Laser Ranging: a Playground for Gravitational Physics\\
Michael Kesden, New Astrophysical Probes of Black Hole Spin\\

Multimessengers from Space\\
(joint with DAP)\\
Elizabeth Hays, Gamma-ray and X-ray Views of the Energetic Sky from a Multimessenger Perspective\\
Eun-Suk Seo, Current and Future Cosmic Ray Observatories in Space\\
James Thorpe, The Science of Gravitational Waves with Space Observatories\\

The GGR contributed sessions are as follows:\\

Alternative Gravity Theories\\

Coalescing Binary Waveforms\\

Quantum Gravity and Cosmology\\

Gravitational Wave Data Analysis\\

Simulations of Binary Neutron Star and Black Hole - Neutron Star Mergers\\

Classical and Semiclassical Gravity\\

Numerical Investigations in Relativity\\

Gravitational Experiments and Instruments\\

Gravitational Wave Observation\\

General Relativity: Mathematical Aspects\\

New Directions in Gravitational Physics\\

Numerical Relativity Methodology\\

Approximation Methods in General Relativity\\

Astrophysical Numerical Relativity Simulations\\

\end{document}